%%%%%%%%%%%%%%%%%%%%%%%%%%%%%%%%%%%%%%%%%%%%%%%%%%%%%%%%%%%%%%%%%%%%%%%%%%%%%5
% 
% set up Font tables
%
%
% Euler Roman (cursive)

 \font\teneurm=eurm10
 \font\seveneurm=eurm8
 \font\fiveeurm=eurm6
 \newfam\eurmfam
 \textfont\eurmfam=\teneurm
 \scriptfont\eurmfam=\seveneurm
 \scriptscriptfont\eurmfam=\fiveeurm
 \def\eurm#1{{\fam\eurmfam\relax#1}}

% Euler Roman math mode

% \textfont1=\teneurm
% \scriptfont1=\seveneurm
% \scriptscriptfont1=\fiveeurm
% \font\ninesym=cmsy9
% \textfont2=\ninesym

 % Euler Script

 \font\teneusm=eusm10
 \font\seveneusm=eusm8
 \font\fiveeusm=eusm6
 \newfam\eusmfam
 \textfont\eusmfam=\teneusm
 \scriptfont\eusmfam=\seveneusm
 \scriptscriptfont\eusmfam=\fiveeusm

 % Euler Extension

 \font\teneuex=euex10
 \font\seveneuex=euex8
 \font\fiveeuex=euex7 at 5pt
 \newfam\euexfam
 \textfont\euexfam=\teneuex
 \scriptfont\euexfam=\seveneuex
 \scriptscriptfont\euexfam=\fiveeuex

% ---------------------------------------------------------------------------
%
% Concrete Roman at 10pt
%
% ---------------------------------------------------------------------------
\font\cninerm=ccr10 \font\csevenrm=ccr8 \font\cfiverm=ccr6
\font\cninei=ccmi10 \font\cseveni=ccmi10 at 8pt \font\cfivei=ccmi10 at 6pt
\font\cninesy=cmsy10 \font\csevensy=cmsy8 \font\cfivesy=cmsy6
%\font\cninebf=cccsc10 \font\csevenbf=cccsc10 at 8pt \font\cfivebf=cccsc10 at 6pt
% cant find mtbx10.mf
%\font\cninebf=mtbx10 \font\csevenbf=mtbx10 at 8pt \font\cfivebf=mtbx10 at 6pt
\font\cninebf=cmbx10 \font\csevenbf=cmbx8 \font\cfivebf=cmbx6
\font\cnineit=ccti10 
\font\cninesl=ccsl10 
\font\cmex=cmex10 \font\sevencmex=cmex10 at 8pt \font\fivecmex=cmex10 at 6pt
\skewchar\cninei='177 \skewchar\cninesy='60

%
% ---------------------------------------------------------------------------
%
% Concrete Roman at 9pt    for figure captions
%
% ---------------------------------------------------------------------------
\font\cnninerm=ccr9 \font\cnsevenrm=ccr7 \font\cnfiverm=ccr5
\font\cnninei=ccmi10  \font\cnseveni=ccmi10 at 7pt \font\cnfivei=ccmi10 at 5pt
\font\cnninesy=cmsy9 \font\cnsevensy=cmsy7 \font\cnfivesy=cmsy5
% cant find mtbx10.mf
%\font\cnninebf=mtbx10 at 9pt \font\cnsevenbf=mtbx10 at 7pt \font\cnfivebf=mtbx10 at 5pt
\font\cnninebf=cmbx9 \font\cnsevenbf=cmbx7 \font\cnfivebf=cmbx5
\font\cnnineit=ccti10 at 9pt 
% \font\cnnineit=ccti10 at 9.5pt \font\cnsevenit=ccti10 at 7pt
% \font\cnninesl=ccsl10 at 9.5pt \font\cnsevensl=ccsl10 at 7pt
\font\cnninesl=ccsl10 at 9pt 
\font\cmex=cmex10 at 9pt \font\sevencmex=cmex10 at 8pt \font\fivecmex=cmex10 at 6pt
\skewchar\cnninei='177 \skewchar\cnninesy='60

%
 % ---------------------------------------------------------------------------
 %
 % lcmss at 8.5pt
 %
 % ---------------------------------------------------------------------------
   
 \font\leninei=cmmi9   
 \font\leninesy=cmsy9

 \font\cmex=cmex10 \font\sevencmex=cmex10 at 8pt \font\fivecmex=cmex10 at 6pt
 \skewchar\leninei='177 \skewchar\leninesy='60

 % ---------------------------------------------------------------------------
 %
 % lcmss at 12pt
 %
 % ---------------------------------------------------------------------------
    
 \font\lptwlvi=cmmi12 at 13pt  
 \font\lptwlvsy=cmsy10 at 13pt

 \font\cmex=cmex10 at 13pt    
 \skewchar\lptwlvi='177 \skewchar\lptwlvsy='60

% ---------------------------------------------------------------------------
%
% lcmss at 8pt          for figure captions
%
% ---------------------------------------------------------------------------
  
\font\lninei=cmmi8   
\font\lninesy=cmsy8

\font\cmex=cmex10 at 9pt \font\sevencmex=cmex10 at 8pt \font\fivecmex=cmex10 at 6pt
\skewchar\lninei='177 \skewchar\lninesy='60

% ---------------------------------------------------------------------------
%
% Computer Modern Roman at 9pt    for figure captions
%
% ---------------------------------------------------------------------------
   
 \font\crninei=ccmi10   
 \font\crninesy=cmsy9  
%\font\crninebf=mtbx10 at 9pt \font\crsevenbf=mtbx10 at 7pt \font\crfivebf=mtbx10 at 5pt
% cant find mtbx10.mf
% \font\crninebf=mtbx10 \font\crsevenbf=mtbx10 \font\crfivebf=mtbx10

 \font\cmex=cmex10 at 9pt \font\sevencmex=cmex10 at 8pt \font\fivecmex=cmex10 at 6pt
 \skewchar\crninei='177 \skewchar\crninesy='60

% ---------------------------------------------------------------------------
% ---------------------------------------------------------------------------
% fonts

% font family switches
%\lcmssten   % switch to lcmss at 8.5pt, very close to roman 10pt
%\lcmss      % switch to lcmss at 8pt                     figures
%\ccrten     % switch to concrete roman 10pt
%\ccrnine    % switch to concrete roman 9pt               figures
%\cmrnine    % switch to computer modern roman 9pt        figures

% ---------------------------------------------------------------------------
% Old font info... data inbedded in the above families
%
% \font\rm=lcmss8 at 8.5pt           	% normal type   %
% \font\it=lcmssi8 at 8.5pt		% italics       %
% \font\bf=lcmssb8 at 8.5pt		% bold type     %
% \font\rm=ccr10			% computer concrete roman  %
% \font\it=ccti10			% concrete italics %
% \font\sl=ccsl10			% concrete slanted %
% \font\bf=cccsc10      		% small caps     %

\font\tt=cmtcsc10            	      % computer script %
		      % typewriter script %

% ---------------------------------------------------------------------------
%           Chapter Heading Font    (chapter bold font)
%			also the header line font, chf

\font\chf=cmcsc10 at 12pt
\font\cbf=cmcsc10 
% if there is a need for subsub chapters, use lcmss8 or ccti

% ---------------------------------------------------------------------------
%                       Figure and Table Caption Font
%% \font\bff=mtbx9
% \let\figfont=\ccrnine
% should replace
% \font\bff=mtbx10 at 9pt
% \font\ff=ccr9 

% \let\figfont=\lcmss
% replaces...
% \font\bff=lcmssb8 				% for fig. cap. titles
% \font\ff=lcmss8   				% for fig. cap. text

% \let\figfont=\ccrnine
\let\figfont=\ccrten

 		% small... for use in the code tables
% \font\sff=ccr9 			% small... for use in the code tables

% ---------------------------------------------------------------------------
%                       Table  Font
% \let\tf=\ccrten
% \let\thf=\ccrten
%\let\tf=\cmrnine
%\let\thf=\cmrnine

% \let\tf=\lcmss
% \let\thf=\lcmssten

\font\thf=cmcsc10
\let\tf=\ccrnine

% ---------------------------------------------------------------------------
%                       special purpose fonts
%
		% for title scale bold font
%\font\bfx=cmbx10 at 11pt		% for section headings
\font\bfx=cmbx12			% for section headings
\font\bf=cmbx10 at 11pt			% for subsection heading

% \font\cmr=cmr9

% \font\sbf=mtbx10

% \font\punk=punk10
% \font\sl=cmti10 scaled 950
% \font\init=yinit
% \font\palin=rpplr at 9pt  
% \font\ccsl=ccsl10
% \font\howl=lcmss8 at 8.5pt           	% howl type   %

% \font\ygoth=ygoth 
% \font\yfrak=yfrak
% \font\yswab=yswab scaled 1050
 
\rm
\topskip= 0.0 cm
  \hsize = 14.413  truecm
%  \vsize = 19.558 truecm
  \vsize = 21.0 truecm
  \hoffset = 0.519 truecm
  \voffset = 0.254 truecm

%
%       choose 9/10ths:                    ytop = 25.1460
%                        with 1cm binding
%               hoffset = 3.059         voffset = 2.794
%               hsize = 14.413          vsize   = 19.558
%
%
% hack to make chapter header number come out right justified
\newdimen\headsize
\headsize=\hsize
\advance\headsize by 4pt
%
% ---------------------------------------------------------------------------
% spacing
%
%	double  spacing
%\baselineskip= 19pt
%\lineskiplimit= 1pt
%\lineskip= 20pt plus 20pt minus 20pt
%\parskip= 5pt
%	single  spacing
\baselineskip= 13pt
\lineskiplimit= 1pt
\lineskip= 14pt plus 2pt minus 2pt
\parskip= 8pt
%
%	figures
\def\figbase{\baselineskip=13pt
\lineskiplimit= 0pt
\lineskip= 1pt plus 2pt minus 2pt}

\def\leaderfill{\leaders\hbox to 1em{\hss.\hss}\hfill}
\def\content#1#2{\line{#1\leaderfill \hbox to 1cm{\hss #2}}}

%-----------------------------------------------------------------------
% counting figures and tables

\countdef\tabno=141
\countdef\figno=142
\countdef\chapno=143
\countdef\sectno=144

%-----------------------------------------------------------------------
% section headings
\def\section#1{
   \global\sectno=0
    \bigskip\smallskip
     \global\advance\chapno by 1 
      \centerline{\bfx \the\chapno. #1} 
       \noindent\hskip 0em
}

% subsection headings
\def\subsection#1{
    \ifnum\sectno>0 \bigskip\smallskip \else {\vskip -0.6cm } \fi
     \global\advance\sectno by 1 
      \line{\bf \the\chapno.\the\sectno\quad #1 \hfill}
       \noindent\hskip 0em
}
%
%-----------------------------------------------------------------------
% references

\def\nhi{\noindent \hangindent=1.0truecm}

%
%-----------------------------------------------------------------------
% math 

% lets define good and bad...
 \def\good{\ifmmode \eurm{G} \else $\eurm{G}$\fi}
 \def\bad{\ifmmode \eurm{B} \else $\eurm{B}$\fi}

%-----------------------------------------------------------------------
% units

\def\deg{\ifmmode ^{\circ} \else $^{\circ}$ \fi}
\def\min{\ifmmode ^{\prime} \else $^{\prime}$ \fi}
\def\s{\ifmmode ^{\prime\prime} \else $^{\prime\prime}$\fi}
\newdimen\sa
\def\sd{\sa=.1em 
           \ifmmode $\rlap{.}$''$\kern -\sa$
           \else \rlap{.}$''$\kern -\sa\fi}

\def\ergsscm{\ifmmode {{ergs}\over{sec-cm^2}} 
		\else {\rm ergs/s/cm$^2$}      \fi}
\def\h{\ifmmode {h_{75}^{-1}} 
		\else  $h_{75}^{-1}$      \fi}

%-----------------------------------------------------------------------
% misc

\def\date{\number\day\ \  \ifcase\month\or
	January\or February\or March\or April\or May\or June\or July\or
	August\or September\or October\or November\or December\fi
	\space\number\year}

\def\1{\phantom{1}}

\def\b-v{\ifmmode (B^\prime-V)\else $(B^\prime-V)$\fi}
\def\v-r{\ifmmode (V-R)\else $(V-R)$\fi}

\def\sbullet{\raise2pt\llap{$\scriptstyle \bullet$\ }}

%-----------------------------------------------------------------------
% figures
% make the figure caption smaller than the text area
% make the font different, smaller, presumably
% advance the fig number, and medskip

% \def\figbox#1{ {\figbase \figfont \leftskip= 1.25cm \rightskip = 1.25cm
% \def\figbox#1{ {\figbase \figfont \leftskip= 1.0cm \rightskip = 1.0cm

\def\figbox#1{ {\figbase \figfont \leftskip= 1.1cm \rightskip = 1.0cm
\noindent\bf Figure \the\figno. \rm #1
\global\advance\figno by 1
\par } 
\medskip
}

%-----------------------------------------------------------------------
% tables
% make the table header in thf font
% make the table font different, smaller, presumably
% advance the tab number, and medskip
% and truy to topinsert

\def\tabbox#1#2{ 
\topinsert{
{\thf 
\centerline{Table \the\tabno.}
\centerline{#1}                    
   }
{\tf 
\input #2
\global\advance\tabno by 1
  }  }
\endinsert
}

\def\tabboxa#1#2{ 
{ {\thf 
\centerline{Table \the\tabno.}
\centerline{#1}                    
   }
{\tf 
\input #2
\global\advance\tabno by 1
  }  }
}

%-----------------------------------------------------------------------
% quotes
%       long quote in a paragraph

%-----------------------------------------------------------------------
% quotes
%       take the quote, redefine line spacing,
%       and put the chapter heading 4cm from top,
%	regardless of how long the quote is.

%-----------------------------------------------------------------------
% quote 2 = Chapter Header Spacing
%	same as above, different spacing

\def\quote2#1{{ \vbox to 2.0cm{
\leftskip=0pt plus 40pc minus \parindent \parfillskip=0pt
\parskip=3pt \baselineskip=3pt 
\vskip 0.75cm
#1
\vfill
}
\vskip -2.0cm
\vskip  2.0cm
}
}

%-----------------------------------------------------------------------
% code box
%       for displaying the code samples in chapter 3
%
% TeX book question 21.3
% \def\boxit#1{\vbox{\hrule\hbox{\vrule\kern0.5cm
% 	\vbox{\kern0.5cm#1\kern0.5cm}\kern0.5cm\vrule}\hrule}}
% My version
\def\boxit#1{\vbox{\hrule\hbox{\kern0.5cm
	\vbox{\kern0.5cm#1\kern0.5cm}\kern0.5cm}\hrule}}

% use it for code block.

\def\codebox#1{{\moveright 1em
\boxit{\hsize=12.5cm \parskip=8pt \baselineskip=20pt \tt
#1
} 
}}

%-----------------------------------------------------------------------
%
% for bibliography
%--------------%
%  REFERENCES  %
%--------------%
\def\jref#1 #2 #3 #4 {{\par\noindent \hangindent=3em \hangafter=1
      #1, {#2}, {#3}, #4\par}}
\def\jreftitle#1 #2 #3 #4 #5 {{\par\noindent \hangindent=3em \hangafter=1
      #1, {``#2,''} {#3}, {#4}, #5\par}}
\def\ref#1{{\par\noindent \hangindent=3em \hangafter=1
      #1\par}}

% set numberings
\figno=1
\tabno=1

% * Fermi's question 
% * type III rate due to non-observance in Galaxy 
% * scaling laws of spirals and ellipticals 
% 	* natural galaxies 
% 	* hubble atlas and rate 
% 	* FP and tully fisher type III rate 
% * evolutionary timescales 
% * hypothesis of natural nemesis 
% 	* gamma-ray burst like analysis 
% * revisiting the rates in a NN world 
% 

\bigskip
\bigskip
\centerline{\chf An Astrophysical Explanation for the Great Silence}
\bigskip
\line{James Annis \hfil}
\line{Experimental Astrophysics Group \hfil}
\line{MS 127, Fermi National Accelerator Laboratory, Batavia, IL 60510, USA \hfil}
\line{annis@fnal.gov}
\bigskip

{\narrower
	An astrophysical model is proposed to answer Fermi's question.
	Gamma-ray bursts have the correct rates of occurrence and plausibly
	the correct energetics to have consequences for the evolution of
	life on a galactic scale. If one assumes that they are in fact lethal to
	land based life throughout the galaxy, 
	one has a mechanism that prevents the rise of intelligence
	until the mean time between bursts is comparable to the
	timescale for the evolution of intelligence. Astrophysically plausible
	models suggest the present mean time between bursts to be $\sim 10^8$ 
	years, and evolutionarily plausible models suggest the rise of
	intelligence takes $\sim 10^8$.
	Hence, this model suggests that the Galaxy is currently undergoing a 
	phase transition between an equilibrium state devoid of intelligent 
	life to
	a different equilibrium state where it is full of intelligent life.
	\bigskip
}

``Where are they?'', Fermi asked [1]. In classic Fermi style, he had worked 
through an order of magnitude calculation capturing the essence of a
problem, in this case the existence of extra-terrestrial civilizations. He had
realized that the
likely time scales of the expansion of a civilization through a 
galaxy are so much shorter than the age of the galaxy itself that, given the existence 
of one intelligent race, one has to wonder where everyone else is. 

Fermi's question is so simple and so powerful it is worth going over it in detail.
First, the Galaxy is roughly $10^{10}$ years old. 
Second, the Galaxy is roughly 100,000 light years across and on average has
a star every light year. The average time it takes a civilization to move
between stars sets it's expansion velocity. At Earth orbital velocities,
$10^{-4}$c, it takes 10,000 years. Technologies
that allow $10^{-1}$c seem feasible, lowering the interstellar time to 10 years.
Take the mean time to be 1000 years, corresponding to $10^{-3}$c, to allow
for the inevitable delay between the colonization of one system and the embarkation
upon a new colonization project.
The time it takes a civilization to colonize from one end of the galaxy to another is 
$t_c \sim r/v \sim 10^8$ years. 
(More careful studies suggest times ranging from 50 million years [2] to 
$10^9$ years [3], showing the power of Fermi's order of magnitude argument.) 
The capstone to Fermi's question is the mismatch between 
the galactic colonization timescale
and the age of the galaxy: $t_g \sim 10^{10} \gg t_c \sim 10^8$.
Once an interstellar 
spacefaring civilization develops, it should sweep across the galaxy like 
wildfire, as viewed on galactic timescales. 

Have they come and gone? 
Clearly the vast majority of humans today are unaware of any contact with 
extra-terrestrials. Just as clearly, the nearest handful of sun-like stars show 
no civilization produced radio emissions. 
% Searches for astronomical objects with the characteristics of Dyson spheres have turned up empty. 
Furthermore, 
the solar system appears primordial in the sense that it does not show 
the effects of the mega-engineering projects seemingly within Mankind's grasp in the 
coming centuries. The Moon appears untouched since the great bombardment
4 billion years ago. Venus and Mars have not been terraformed. 
The asteroids show every evidence of being in the orbits they formed in. 
There is no evidence of past or present presence of extra-terrestrial 
civilizations in the solar system. Where are they? 

Brin [4] named this the Great Silence, and cataloged explanations for this 
silence. Here I wish to add a new one: an astrophysical reason for the late 
development of intelligent life in the galaxy. 

% The paradox revolves around time scales. First, the universe is
% roughly 15 billion years old. Second, the Galaxy, and galaxies in
% general are roughly 100,000 light years in diameter and on
% average have a star every light year or so. The argument:
% how long does it take to move between a pair of stars? At 0.1c,
% 10 years. At Earth orbital velocities, 0.0001c, 10,000 years.
% Take 1,000 years, on average, and leave open the question
% as to whether this is due to low space velocity or to time between
% successive colonization events. At $10^3$ years between stars,
% it takes 100 million years to go from one side of the galaxy to the other.
% A more careful study suggests perhaps 60 million years to fill the entire
% galaxy with colonized planets [2], but then, a different model yields
% longer timescales [3].
% The capstone to Fermi's question is then the mismatch between
% the age of the Galaxy, $10\times10^{9}$ years, and the galactic colonization
% timescale, $0.1 \times10^9$ years. Once an interstellar space faring
% civilization developed, it should sweep across the Galaxy like
% wildfire when viewed on galactic timescales. Where are they?

% Fermi's question makes just two assumptions: that
% civilizations are long lived and that travel between stars is possible.
% As one way travel between stars is easily within the conceptual
% horizons of current science, Fermi's paradox rapidly turns into
% questions about civilizations.

\bigskip
\centerline{\cbf An Astrophysical Explanation}
Gamma-ray bursts are 10 second flashes of ${1\over2}$ Mev photons
occurring 300 times a year at positions isotropically distributed on the sky.
It has been shown that they are at cosmological distances [5]. 
The energy output during a burst is enormous, $10^{52}$ ergs over a few seconds. 
This is comparable to that released by supernovae, but differs in two
important ways.
First, the bulk of the energy released by a supernova escapes in the
form of neutrinos. We know too little about gamma-ray bursts to know
if they also release neutrinos.
Second, the small amount of energy that escapes from supernovae
as photons must work its way out of an opaque fireball, a process that takes
place over days. In gamma-ray bursts the photons are released in seconds.

If a gamma-ray burst occurs anywhere inside the observable
universe, we see it. The observed rate  of 300 times a year
gives over a $10^6$ years $3\times10^8$ bursts. 
The observable universe contains about $10^9$ galaxies,
so the data are consistent with each galaxy having a gamma-ray burst
every $3\times10^6$ years.  This is intriguingly close to evolutionary time scales.

The rate of gamma-ray bursts almost certainly was higher in the past than
in the present. There are many astrophysical
reasons to expect the number to decline with time, both
empirical and model dependent.
For example, the leading contender for the cause of gamma-ray burst 
is colliding neutron stars. 
These would have been born in binary systems and fairly rapidly spiral
inward. Their numbers reflect the star formation history
of the universe, which peaked 10 billion years ago and has declined since.

We can model this with a per galaxy
gamma-ray burst rate that falls exponentially with time:
$$r_\gamma(t) = r_\gamma(0) e^{-t/\tau} \,\, ,$$
where $\tau$ is the decay time constant. Take a plausible
$\tau = 2.5$ billion years. Given the mean rate of
 $\bar r^{-1} = 3\times10^6$ years we know $r_\gamma(0)$,
 and can predict the rate at any time $t$. 
At present $t=15$ billion years, so the current rate is
$t_\gamma = r^{-1}_\gamma({\rm now}) \approx 220$ million years.

The crux of the explanation is this: assume that each
gamma-ray bursts is a mass extinction event on a galactic scale.
The energetics of gamma ray bursts are such that if one
went off in the galactic center (and absorption by 
dust was not a factor) then we here two-thirds the way out the galactic
disk would receive, over a few seconds, and all in 300 keV $\gamma$ rays,
the equivalent of 1/10 the solar flux. 
The assumption of lethality is an assumption that there exists
some mechanism that would translate the energy in photons to
lethal radiation at the planetary surface. 
The atmosphere prevents the $\gamma$-ray photons from being directly
lethal, but bursts may damage the atmosphere.
Thorsett [6] catalogs the effects of nearby gamma ray bursts on the
atmosphere: the strongest possibility is the disruption of the ozone layer,
where a $50\%$ reduction in ozone results in a factor of 50 increase
in the 295 nm solar  UV flux most disruptive to protein structure.
Uv absorption here is highly lethal to all known forms
of cellular activity. There are other possible mechanisms.
Here I will explicitly assume that gamma-ray bursts
are lethal.

The gamma-ray burst model is therefore one where galactic scale
mass-extinctions occur often.
Ten billion years ago, the rate was quite high, perhaps
every 3 million years. Over time this rate slows down and now
the rate is perhaps once every 220 million years. Given
the premise of this model, the last such burst in our Galaxy was 
before the solid surface of the Earth was covered with life,
270 million years ago.
 These bursts are not likely to be lethal to
an advanced civilization, so their effectiveness at preventing
the Galaxy from being colonized lies in their effectiveness are
preventing intelligent life from evolving in the first place.
We will need a timescale for the evolution of intelligence.

\bigskip
\centerline{\cbf Evolutionary Timescales}
There are many evolutionary timescales evident on Earth.
It took less than $1$ billion years for life to form, and $3$ billion
before it moved beyond simple multi-cellular creatures. 
The last ${1\over2}$ billion years saw the rise of 
fish, land animals, and intelligence. Of this,
300 million years were dominated by fish, 200 million years were 
dominated by dinosaurs, and 70 million by mammals. 

% Life on land is perhaps 270 million years old. The gamma-ray burst
% model need not have anything to do with the mass extinction events
% since then: comets and climate change are more than satisfactory.
% The gamma-ray burst model does explain the 4.3 billion years before
% life took over the land: every time it tried before that it was annihilated
% by a gamma-ray burst.

But the relevant timescale is the rise of intelligent life. The rise need not
start from unicellular life. The gamma-ray burst model 
predicts lethal events occurring on 1-100 million year timescales. 
It is built into the assumptions of the model that most life on land cannot 
survive through a burst. What matters is the timescale for land animals
to develop intelligence: while this is long compared to 
the mean time between bursts in a galaxy, intelligence cannot form.

One could take the full 400 million years that life has been on
land as the timescale, but this is an overestimate and is misleading.
 Fish have had 500 million years to develop language, tool, and math using
intelligence and never have; dinosaurs were an immensely successful form
that had 200 million years filling every possible land-based niche, and they
did not rise to intelligence. Mammals spent 100 million years
competing with dinosaurs and did not rise to intelligence, but allowed
to test the limits of their form, it took just 70 million
years to do so.

The issue here is probably the complexity of the organisms. Bonner [7] argues that 
there has clearly been an increase in complexity of organisms over time. He 
uses raw size as a proxy for complexity.
The argument is that the more complex a 
phyla or class is, the larger the organism can be built on that model. Not that 
the largest creatures are the most successful: far from it. These creatures are 
on the margin and do not survive well mass extinction events. But the smaller 
members of the more complex phyla or classes tend to survive and dominate the 
next era. 
It is demonstrably true that the largest creature in both the seas and on the 
land has been getting larger. 

An interesting speculation is that evolution can not produce intelligence 
until it has a complex enough base class to start with. The data tend to 
support this, as we can show using brain mass. 
There is a scaling relation between body and brain: brain mass 
$\propto$ body mass$^{3/4}$ which one  can see in figure 1. 
The significant point is that the earlier evolutionary classes of
 fish, reptiles, and 
dinosaurs lie on one scaling relation while the later classes of 
birds and mammals lie
on a second. These relations have the same slope but differ in zeropoint, with 
the
later classes having a factor of 10 larger brain mass for a given body mass.
Not surprisingly, to my mind, birds and mammals are in general smarter and exhibit more complex 
behavior than either fish or reptiles. 
Primates are on the high side of the higher scaling relation. Humans and porpoises 
scatter off the high side. Suggesting that land base is important,
as it was the land based humans that developed civilization.

\topinsert
\vbox to 3.4in{\hfil}
\includegraphics{data/brain.ps}
\figbox{
The brain mass -- body mass diagram. The scaling relation between brain and body is
m$_{brain} \propto$ m$_{body}^{3/4}$. Different classes of vertebraes have different
relation zeropoints, with the later classes of birds and mammals having a factor of 10
larger brain mass then the earlier classes of fish and reptiles at a given body mass.
The solid  lines are least square fits to the early and late classes; both lines have
similar slopes. The roughly 200 million years separating the origins of the classes in
the two relations is taken as the minimum time scale for the evolution of classes
complex enough to support high intelligence. The data are from Bonner [6].
}
\endinsert

It took perhaps 200 million years of land-based evolution to produce
creatures on the upper brain mass -- body mass relation. Once a mass extinction
event eliminated the dinosaurs, releasing the evolutionary niches they controlled,
evolution could experiment with the more complex classes of phyla chordate.
In roughly 100 million years, intelligence formed in the mammals.
Evoluton had far more time to explore the limits of
the classes on the lower complexity scale, but
their limits were short of intelligence.

We have therefore argued that there are only two evolutionary timescales of 
relevance. The first is the $t \sim 10^8$ 
evolution takes 
to develop sufficiently complex base classes of land-dwelling creatures. The 
second is the $t\sim 10^8$ year time scale for intelligence to develop given 
sufficiently complex base classes.

\bigskip
\bigskip
\centerline{\cbf A Phase Transition}
Astronomers have long since learned the power of the Copernican principle:
humans do not occupy a special place in the universe. Arguments
such as ``intelligent life is incredibly improbably, hence we are
likely alone in the universe'' therefore strike astronomers as  themselves
incredibly improbably.
The fact that humans stand on the Earth and look out into the universe
suggests that others are doing the same or have done the same. Yet
they are not here, now.

Brin [4] is responsible for pointing out the power of the idea of equilibrium 
in this context. It is usually fruitful, when considering some new system,
to assume that it is in equilibrium. Most often 
changes can be considered a perturbation on a quasi-equilibrium state. 
This is true for the work done by an expanding gas and it is true for
when considering a galaxy on billion year timescales.
Most responses to Fermi's question, though, propose states of profound
non-equilibrium. Both ``they haven't had time to reach here'' and
``we are first and alone'' (with its implication of our 
standing on the brink of colonizing the galaxy) imply situations far from
equilibrium without any explanation for this better than random chance (or, equivalently,
an appeal to the anthropic principle's ``just so'' story).
The gamma-ray burst model presented here does suggest an explanation: the Galaxy
is presently undergoing a phase transition.

The concept of a phases transition provides a
clear and powerful model for the transition from one equilibrium 
state to another.  Examples include the transformation of water into ice as the
temperature drops, and the transition from an optically
thick to an optically thin universe as electrons first combine with protons
to form hydrogen, again as the temperature dropped. 
For our purposes, the ``phase transition'' can be defined as 
the lowering of a suppressive force below some threshold, past which a 
previously forbidden process becomes allowed. 

It is possible that the rise of intelligence in the universe is a 
phase transition, and that now, all over the galaxy, there are races
standing up and looking out into space. Given the 10 billion year age
of the galaxy and 0.1 billion year time scale for galactic colonization,
this would look like a remarkable coincidence, a remarkable feat of
synchronization. But that is the nature of a phase transition:
once a process is allowed it happens everywhere and rapidly.

The important element in this argument is the nature of 
the suppressive force, which requires a process
able to suppress the rise of civilizations throughout the Galaxy,
and perhaps the universe. Gamma-ray bursts may play such a role.

The timescales are on the same order.
The toy model for the evolution of gamma-ray bursts
suggests $t_\gamma \approx 2\times10^8$. Our arguments
about evolution suggest a timescale of $10^8$ years 
for the rise of organisms complex
enough to support intelligence, and another $10^8$ years
for the rise of intelligence given those organisms,
suggesting an evolutionary timescale of $t_{evo} \approx 2\times10^8$ years.
We have the situation of $t_\gamma \approx t_{evo}$, and we therefore can
expect a phase transition to occur.
(Interestingly enough, the galactic colnization timescale is also $10^8$ years.)

The model essentially resets the available time for the rise of intelligent 
life to zero each time a gamma-ray burst occurs. It therefore answers Fermi's
question  with ``They haven't had time to get here yet''. More importantly, it defeats 
his analysis by changing one of his timescales: the time available for the rise 
of intelligence is changed from the age of the galaxy, $10^{10}$ years, down to 
the time since the last gamma-ray burst, $10^8$ years. Since that timescale is 
comparable to that needed for 
 a) the rise of intelligence, and 
 b) the colonization of the galaxy, 
 Fermi's question loses it's power. 

% The clock is different for each galaxy. Presumably it is 
% reset to zero more often in galaxies with more recent large scale star 
% formation, and with galaxies with overall more stars. 
% 

\bigskip
\centerline{\cbf Implications}
Mass extinction events on Earth were for long thought to be totally
due to internal events: atmospheric transformation, climate change,
or increased competition from other organisms. In the 1980's it became
clear that the Earth is not a closed box, and the other Solar System
bodies, namely comets and asteroids, could cause mass extinction events.
This paper points out that the Solar System itself is not a closed
box, and in principle mass extinction events could be caused by
galactic events, and could be on galactic scales. We were never
promised that the Galaxy is a safe place to live.

If the lethal gamma-ray burst model is correct, the current mean time between 
galactic mass extinction events is 200 million years. If our speculations on
evolution and complexity are on the right track, then once organisms are able 
to survive on land, the time scale for the rise of intelligence is 200 
million years. The timescales match. Thus the non-equilibrium 
condition of at least one species looking out into the galaxy and thinking 
about interstellar travel, yet not living in a galaxy already packed full of 
intelligent life is explained as the onset of a phase transition. A previously 
forbidden configuration is now allowed. It is likely that intelligent life has 
recently sprouted up many at places in the Galaxy, and that at least a few are 
busily engaged in spreading. In another $10^8$ years, a new equilibrium state 
will emerge, where the galaxy is completely filled with intelligent life.

% This 200 million years makes the limit placed by the fundamental
% plane relations to be $r^{-1} = 4.3$ Gyr. This is getting
% to the implausible stage again, but the model is breaking down
% because the time available is not large compared to the time
% for type III formation. For the current rate of 200 million years,
% half of all galaxies should not have experienced a gamma ray burst in
% the last 200 million years. But for type III we need 400 million years
% from insensisence to Dyson spheres, and perhaps $10\%$ of galaxies
% have had this, and the type III rate is  $r^{-1} = 430$ Myr.
% But then, make the sample a factor of 10 bigger, and the rates
% go down into meaningless again.
% 
% One can slide a bit by allowing humans to be a 3 sigma long
% interval between galactic die offs.
% 
% This side track does not help us directly, as we already allowed
% 5 Gyrs of the universe's 15 Gyr for sapient life to develop.
% 

\bigskip
\bigskip
\centerline{\cbf References}
\medskip

\nhi 1. Eric Jones, ``Fermi's Question'', {\it Interstellar Migration and the
	Human Experience}, eds. B.R. Finney and E.M Jones, University of
	California Press, Berkeley, p 298, (1985).

\nhi 2. Eric Jones, ``Colonization of the Galaxy'', {\it Icarus}, 28, 421 (1976).

\nhi 3. William Newman and Carl Sagan, ``Galactic Civilization: Population
	Dynamics and Interstellar Diffusion'', {\it Icarus}, 46, 293 (1981).

\nhi 4. David Brin, ``The `Great Silence': the Controversy Concerning
	Extraterrestrial Intelligent Life'', {\it QJRAS} 24, 283 (1983).

\nhi 5. The literature on gamma-ray bursts is large and growing.
	An accessible start can be found in Martin Rees,
	``Gamma-ray Bursts: Challenges to Relativistic Astrophysics'',
	{\it Proceedings of the 18th Texas Conference}. Other papers
	in that proceedings go much farther into the details.
	The recent evidence that gamma-ray bursts are cosmological is
	reported in M. Metzger et al, {\it Nature} 387, 878 (1997).

\nhi 6. S.E. Thorsett, ``Terrestrial Implications of Cosmological Gamma-ray
	Burst Models'', {\it The Astrophysical Journal}, 444, L53 (1995).

\nhi 7. John Tyler Bonner, {\it The Evolution of Complexity by Means of
	Natural Selection}, Princeton University Press, Princeton (1988).

\vfill\eject
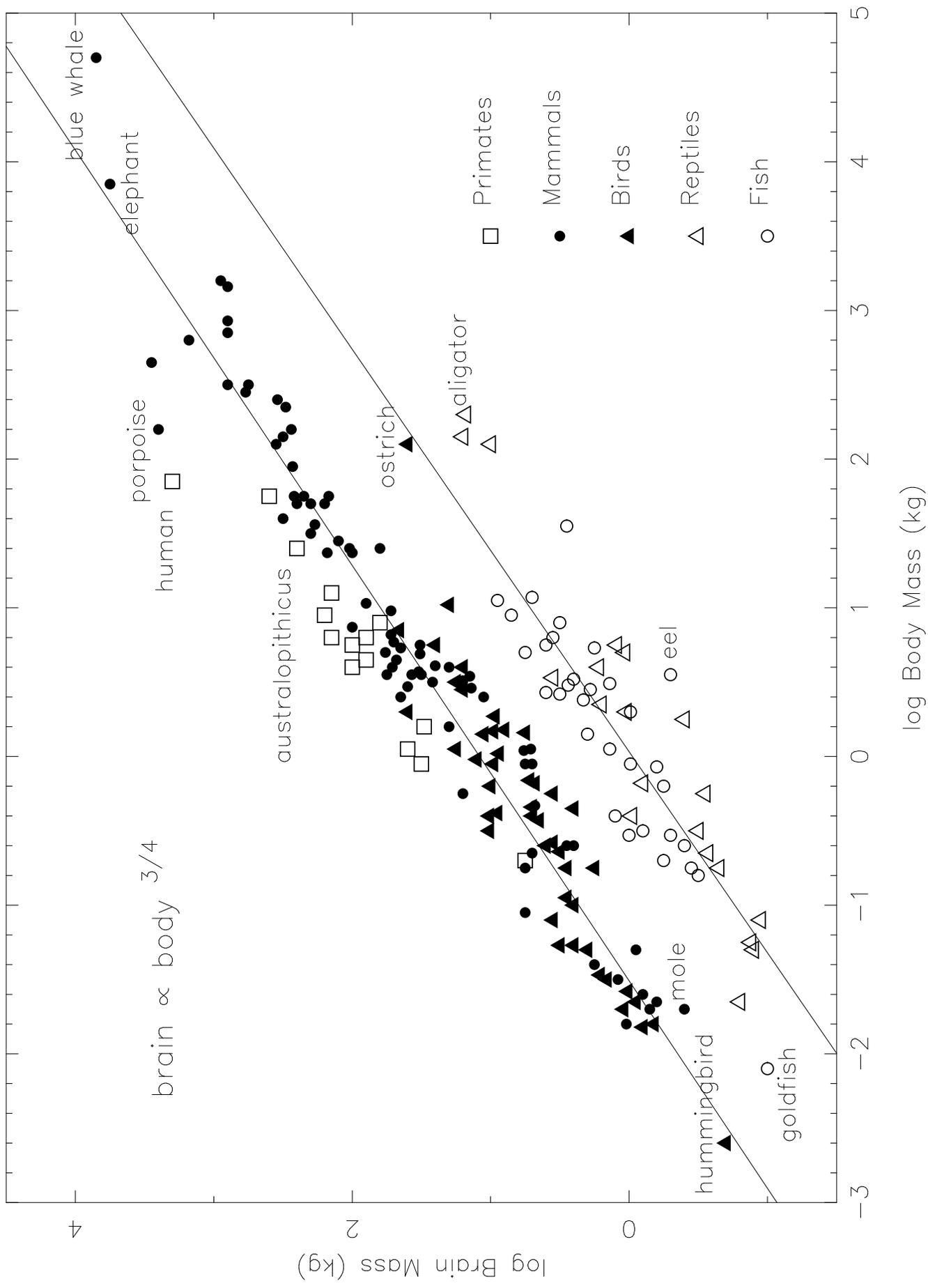
\bye